\documentclass[twoside]{article}
\usepackage{qic,amssymb}

\newtheorem{fact}{Fact}[section]    

\newcommand{\ket}[1]{| #1 \rangle}
\newcommand{\ketbra}[1]{| #1 \rangle \langle #1 |}

\newcommand{\Tr}{{\mathsf{Tr}}}

\newcommand{\defeq}{\stackrel{\Delta}{=}}
\newcommand{\C}{{\mathbb C}}
\newcommand{\R}{{\mathbb R}}

\newcommand{\E}{{{\mathbb E}}}

\newcommand{\cS}{{\cal S}}
\newcommand{\cK}{{\cal K}}
\newcommand{\cH}{{\cal H}}

\begin{document}

\centerline{\bf COMMUNICATION COMPLEXITY OF REMOTE STATE PREPARATION }
\vspace*{0.035truein}
\centerline{\bf WITH ENTANGLEMENT}
\vspace*{0.37truein}
\centerline{\footnotesize Rahul Jain \footnote{{\sf rahulj@cs.berkeley.edu},
2020 Bancroft Way, Berkeley, CA, 94704.}
\footnote{
This work was supported by an Army Research Office (ARO), North
California,  grant number DAAD 19-03-1-00082.
}}
\vspace*{0.015truein}
\centerline{\footnotesize\it Computer Science, University of California,
} 
\baselineskip=10pt
\centerline{\footnotesize\it Berkeley, California, 94720,
USA}
\date{}
\vspace*{0.21truein}
\abstracts{
We consider the problem of {\em remote state preparation} recently
studied in several papers. We study the {\em communication complexity}
of this problem, in the presence of entanglement and in the scenario of single use of the channel. }{}{}

\section{Introduction}
The remote state preparation problem has been studied in in several
papers in recent times, see for example,
~\cite{lo:remote},~\cite{shor:remote},~\cite{benn:remote},~\cite{pati:remote},
~\cite{zhang:remote}.  We define the problem below. Let $X$ be a
set. Let $\cS(\cK)$ be the set of quantum states in the Hilbert space
$\cK$. Let an {\em encoding} $E: X
\mapsto \cS(\cK)$ be a function from $X$ to $\cS(\cK)$. The remote
state preparation, $RSP(X,E,\epsilon)$ problem is as follows:
\begin{definition}[Remote state preparation]
Let Alice, who knows the function $E$, get an input $x \in X$. Alice
and Bob are required to communicate and at the end of the
communication Bob should have a quantum state $\rho_x$ such that
$F(\rho_x, E(x)) \geq 1 - \epsilon$, for some $0\leq \epsilon <
1$. Alice and Bob may start with some prior entanglement between
them.
\end{definition}

In several papers in the remote state preparation problem, Alice
instead of $x$ is given a {\em description} of the state $\rho_x$. We
assume in this work that the description is given in the form of the
element $x$ of $X$. In~\cite{shor:remote}, Bennett, Hayden, Leung,
Shor and Winter studied the trade-off between the {\em rate} of
communication and the rate of entanglement used. In some other papers
like~\cite{benn:remote,pati:remote, lo:remote} the rate of
communication required for this problem was studied with free use of
the entanglement. In most of these earlier works the problem was
studied in the {\em asymptotic setting} where multiple uses of the
communication channel between Alice and Bob were considered. We study
the communication complexity (i.e. the best possible
communication with which a given problem $RSP(X,E,\epsilon)$ can be
solved) of this problem in the scenario of single use of the
channel. By $Q^{pub}(RSP(X,E,\epsilon))$ we denote the
communication complexity, with prior entanglement, of
$RSP(X,E,\epsilon)$. Please note that we are concerned with the total
communication and not the rate as in the earlier papers. Also in this work
we are not concerned with the amount of entanglement used.

We consider a notion of {\em maximum possible information} $T(E)$ in
an encoding $E$ and show that, in the presence of entanglement, the
communication required for $RSP(X,E,0)$, is at least $T(E)/2$ and
$RSP(X,E,\epsilon)$ can be solved with communication at most
$\frac{8}{\epsilon^2}(4T(E) + 7)$. Thus $T(E)$ almost tightly
characterizes the communication complexity of the remote state
preparation problem. It was pointed to us by an anonymous referee that
in one of the main results in [BHL+05], the authors have also
emphasized the role of T(E) for remote state preparation: there it is
shown that the communication cost of preparing tensor products of $n
\rightarrow \infty$ many pure states from the family $E$ of pure states (that would be the
family $E^{\otimes n}$) with an allowable constant fidelity loss is 
$\frac{n}{2} T(E) + o(n)$, so that the lower bound is indeed tight in
this asymptotic setting.

There is an interesting point of note here. In earlier works since the
problem was that of determining the rate of communication and rate of
entanglement etc. in the asymptotic setting, the exact multiplicative
constant in the rate was also important. Since we are concerned
with the total communication in single use of the channel, the problem
of identifying the best communication for a given $RSP(X,E,\epsilon)$
even up to constants is non-trivial. It is easy to see that in
specific cases like when $T(E)=
\log d$, where $d$ is the dimension of $\cK$, or when $T(E) =0$, that
$RSP(X,E,\epsilon)$ can be solved with communication which is like
$T(E)$ up to constants. But for general values of $T(E)$ this problem
is non-trivial.

\section{Preliminaries}
In this section we give a few definitions and state some facts that we
will use later.  

Given a joint quantum system $AB$, the mutual information between them
is defined as $I(A:B) = S(A) + S(B) - S(AB)$, where $S(A)$ is the
von-Neumann entropy of the system $A$.  Given two quantum states, $\rho,
\sigma$ the relative entropy between them is defined as $S(\rho ||
\sigma) \defeq \Tr \rho (\log \rho - \log \sigma)$.  
Let $X$ be a finite set (below we always assume that $X$ is a finite
set) and let $E: x \in X \mapsto \rho_x$ be an encoding over $X$. For
a probability distribution $\mu = \{p_x\}$ over $X$ let $X_{\mu}(E)$ be the
bipartite state $\E_{\mu}[\ketbra{x}\otimes \rho_x] 
\defeq \sum_{x \in X}  p_x \ketbra{x} \otimes \rho_x$. Below
$\E_{\mu}[.]$ always stands for probability average (expectation)
under distribution $\mu$ of the corresponding quantity. Please note
the difference in the font with the notation for an encoding, which is
represented by an $'E'$. Let $I^X_{\mu}(E)$ be the mutual information
between the two systems in $X_{\mu}(E)$. When the underlying set
$X$ is clear we omit the superscript. Let $\rho_{\mu}
\defeq \E_{\mu}[\rho_x]$. We note that in this case from definitions
$I^X_{\mu}(E) = \E_{\mu}[S(\rho_x||\rho_{\mu})]$.

\begin{definition}(Maximum possible information)
{\em Maximum possible information} in an encoding $E: X \mapsto
\cS(\cK)$ is defined as $T_X(E) \defeq \max_{\mu} I^X_{\mu}(E)$. When
the underlying set $X$ is clear we omit the subscript. It is 
easily seen that if $d$ is the dimension of $\cK$ then $T(E)\leq \log
d$. 
\end{definition}

We use the following information-theoretic result called
the {\em substate theorem} due to Jain, Radhakrishnan, and
Sen~\cite{jain:substate}.
\begin{fact}[Substate theorem, \cite{jain:substate}]
\label{fact:substate}
Let $\cH, \cK$ be two finite dimensional Hilbert spaces and 
$\dim(\cK) \geq \dim(\cH)$. Let $\C^2$ denote the two dimensional
complex Hilbert space.  Let $\rho, \sigma$ be density matrices in
$\cH$ such that $S(\rho \| \sigma) < \infty$.  Let 
$\ket{\overline{\rho}}$ be a
purification of $\rho$ in $\cH \otimes \cK$. Then, for $r > 1$, there
exist pure states $\ket{\phi}, \ket{\theta} \in \cH
\otimes \cK$ and 
$\ket{\overline{\sigma}} \in \cH \otimes \cK \otimes \C^2$, 
depending on $r$, such
that $\ket{\overline{\sigma}}$ is a purification of $\sigma$ and
$F(\ketbra{\overline{\rho}}, \ketbra{\phi}) \geq 1- 
 \frac{1}{\sqrt{r}}$, where
\begin{displaymath}
\ket{\overline{\sigma}} \defeq
\sqrt{\frac{r-1}{r 2^{r k}}} \, \ket{\phi}\ket{1} +
\sqrt{1 - \frac{r-1}{r 2^{r k}}} \, \ket{\theta}\ket{0}
\end{displaymath}
and $k \defeq 8 S(\rho \| \sigma) + 14$. 
\end{fact}

The following fact can be found in Cleve et al~\cite{cleve:ip}.
\begin{fact}
\label{fact:cleve}
Let Alice have a classical random variable $Z$. Suppose Alice and Bob
share a prior entanglement independent of $Z$. Initially Bob's qubits
have no information about $Z$.  Now let Alice and Bob run a quantum
communication protocol, at the end of which Bob's qubits possess $m$
bits of information about $Z$. Then, Alice has to send at least $m/2$
qubits to Bob.
\end{fact}

We will require the following minimax theorem from game
theory(see~\cite{osborne:gametheory}). 
\begin{fact}
\label{fact:minimax}
Let $A_1,A_2$ be non-empty, either finite or convex and compact
subsets of $\R^n$. Let $u: A_1 \times A_2 \mapsto \R$ be a
continuous function. Let $\mu_1, \mu_2$ be distributions on $A_1$ and
$A_2$ respectively. Then,
\begin{displaymath}
\min_{\mu_1}\, \max_{a_2\in A_2} \E_{\mu_1}[u(a_1,a_2)] 
= \max_{\mu_2}\, \min_{a_1 \in A_1} \E_{\mu_2}[u(a_1,a_2)]
\end{displaymath}
\end{fact}

We will also require the following Local transition
theorem~\cite{mayer:imp, hklo:bitcomm1, hklo:bitcomm2}. 
\begin{theorem}
Let $\rho$ be a quantum state in $\cK$. Let $\ket{\phi_1}$ and $
\ket{\phi_2}$ be two purification of $\rho$ in $\cH \otimes \cK$. Then
there is a local unitary transformation $U$ acting on $\cH$ such that
$(U \otimes I) \ket{\phi_1} = \ket{\phi_2}$. 
\end{theorem}

\section{Communication bounds}

The following lemma states the communication lower bound.
\begin{lemma}
Let $E : x \mapsto \rho_x$ be an encoding, then $Q^{pub}(RSP(X,E,0))
\geq T(E)/2$.
\end{lemma}

\proof{
Let $T(E) = c$. Let $\mu$ be the distribution on $X$ such that
$I_{\mu}(E) = c$. Consider the random variable $Z$ taking values in
$X$ with distribution $\mu$. Let Alice be given inputs according to
$\mu$. We know that after the remote state preparation protocol mutual
information between $Z$ and the qubits of Bob, where the state is
created, is $c$. Hence by fact~\ref{fact:cleve} at least
$c/2$ qubits must be communicated by Alice to Bob.}

\noindent
{\bf Remark:} As suggested by an anonymous referee we out point here that
the above lemma is not robust for positive $\epsilon$. This is because
after allowing for a small error $T(E')$ may be smaller than
$T(E)$ by up to order  $\epsilon \log d + \epsilon \log \epsilon$, where $E'$ is
the new encoding obtained by allowing the positive error
$\epsilon$. This follows from Fannes inequality~\cite{fannes:ineq}.

On the other hand we show the following upper bound on the
communication required to solve the problem.
\begin{theorem}
Let $E : x \mapsto \rho_x$ be an encoding and $0 < \epsilon <1 $
be a constant, then $Q^{pub}(RSP(X,E,\epsilon)) \leq \frac{8}{\epsilon^2}(4T(E) + 7)$.
\end{theorem}

\proof{
We first show the following key lemma.
\begin{lemma}
\label{lem:SleqT}
Let $E : x \mapsto \rho_x$ be an encoding. There exists a
distribution $\mu$ such that $$\forall x \in X, S(\rho_x|| \rho_{\mu})
\leq T(E) $$
\end{lemma}

\proof{
Let $A_1$ be the set of all distribution on the set $X$. Let $A_2$ be
the set $X$ itself. The function $u: A_1
\times A_2 \mapsto \R$ be such that $u(\mu,x) = S(\rho_x
|| \rho_{\mu})$. The conditions of Fact~\ref{fact:minimax} are
satisfied and therefore we have:
\begin{eqnarray}
 \min_{\mu} \max_x S(\rho_x|| \rho_{\mu}) & \leq & \min_{\mu*:
\mbox{distribution over distributions $\mu$}} \max_{x} \E_{\mu*}[S(\rho_x || \rho_{\mu})]  
\\
& = & \max_{\lambda : \mbox{distribution over $X$}}
\min_{\mu} \E_{\lambda}[S(\rho_x || \rho_{\mu})]
\\
& \leq & \max_{\lambda}
\E_{\lambda} [ S(\rho_x|| \rho_{\lambda})]  \\
& = & \max_{\lambda}  I_{\lambda}(E) = T(E)
\end{eqnarray}
Inequality (1) follows since relative entropy is jointly convex
in its arguments. Equality (2) is from Fact~\ref{fact:minimax}.
}

Let $T(E)=c$, then from lemma~\ref{lem:SleqT} we get a
distribution $\mu$ on $X$ such that $ \forall x, S(\rho_x ||
\rho_{\mu} ) \leq c$. Let Alice and Bob start with $2^{rk}$ ($r = 4/\epsilon^2, k = 8c + 14$) copies of some
purification $\ket{\psi}$ of $\rho_{\mu}$ with the purification part
being with Alice and $\rho_{\mu}$ with Bob in each of the copies of
$\ket{\psi}$.  Let us invoke  Fact~\ref{fact:substate} with $\rho
\defeq \rho_x, \sigma \defeq \rho_{\mu}$ and $\ket{\overline{\rho}}$
being any purification of $\rho_x$. Let $\ket{\psi_x}$ be the
purification of $\rho_{\mu}$ obtained from Fact~\ref{fact:substate}
corresponding to $\ket{\overline{\sigma}}$. Since the reduced quantum
state on Bob's part in both $\ket{\psi_x}$ and $\ket{\psi}$ is the
same, from local transition theorem, there exists a transformation
acting only in Alice's side which takes $\ket{\psi}$ to
$\ket{\psi_x}$. Alice on input $x$, transforms each $\ket{\psi}$ to
$\ket{\psi_x}$ and measures the first bit. If she obtains 1 in any
copy of $\ket{\psi_x}$ she communicates the number of that copy to
Bob. It is easily seen that the communication from Alice is
at most $rk = \frac{8}{\epsilon^2}(4c + 7)$. Also since $\Pr(\mbox{Alice observes $1$}) = \frac{r-1}{r 2^{r
k}}$, and Alice makes $2^{r k}$ tries she succeeds
with probability at least $1 - 1/r$. In case she succeeds, let the
state with Bob in which Alice succeeds be $\rho_x'$. From
Fact~\ref{fact:substate}, $F(\rho_x', \rho_x) \geq 1 - 1/\sqrt{r}$.
So for the final state $\tilde{\rho_x}$ produced with Bob, it follows
from concavity of fidelity that $F(\tilde{\rho_x}, \rho_x) \geq 1 - 2/\sqrt{r} = 1 - \epsilon$.
}
\\ 
\noindent \\
{\bf Remarks:} 
\begin{enumerate}
\item 
Given an encoding $E:x \mapsto
\rho_x$, a small constant $\epsilon$ and states
$\rho_x'$ such that $F(\rho_x', \rho_x) \geq 1 - \epsilon $, let a
{\em perturbed encoding} $E'$ be, $E': x \mapsto \rho_x'$. It is quite
possible that $T(E')$ is much less than $T(E)$ as allowed by
Fannes bound. In such a case communication can be reduced a lot by
running the above protocol for $E'$ instead of $E$ since we are ready
to tolerate constant fidelity loss anyway.
\item 
One can consider the classical version of the remote state generation problem 
in which the encoding considered is a mapping from $X$ to the set of classical
distributions on some set. On input $ x \in X$ to Alice, they are
required to communicate, at the end of which Bob is required to sample
from a distribution close to $E(x)$. The same communication bounds
apply for this problem as well.  
\end{enumerate}

\section{Conclusions}
The protocol for the upper bound, mentioned in this paper uses a large
amount of entanglement. It will be interesting to see if it can be
reduced or even eliminated if possible. Also it will be interesting to
get entanglement-communication trade-offs for this problem as opposed
to the trade-offs in the rates of entanglement and communication
mentioned in some of the earlier works.

\section{Acknowledgment}
We thank Rohit Khandekar, Pranab Sen, Julia Kempe and Jaikumar
Radhakrishnan for useful discussions and for pointing out useful
references. We also thank anonymous referees for useful suggestions
and comments.

\nonumsection{References}

\newcommand{\etalchar}[1]{$^{#1}$}


\begin{thebibliography}{CvDNT98}

\bibitem{lo:remote}
H.-K. Lo (2000), {\it Classical communication cost in distributed
quantum information processing - a generalization of quantum
communication complexity}, Phys. Rev. A, 62.

\bibitem{shor:remote}
C.H. Bennett, P.~Hayden, W.~Leung, P.W. Shor, and A.~Winter (2005),
{\it Remote preparation of quantum states}, IEEE transaction of
information theory, 51, pp 56-74.

\bibitem{benn:remote}
C.H. Bennett, D.P. DiVincenzo, P.W. Shor, J.A. Smolin, B.M. Terhal, and W.K.
  Wootters (2001), {\it Remote state preparation}, Phys. Rev. Lett, 87.

\bibitem{pati:remote}
A.K. Pati (2001), {\it Minimum classical bit for remote preparation and measurement of a
  qubit}, Phys. Rev. A, 63.

\bibitem{zhang:remote}
B.~Zeng and P.~Zhang (2002), {\it Remote-state preparation in higher dimension and the parallelizable
  manifold $s^{n-1}$}, Phys. Rev. A, 65.

\bibitem{jain:substate}
R.~Jain, J.~Radhakrishnan, and P.~Sen (2002), {\it Privacy and interaction in quantum communication complexity and a
  theorem about the relative entropy of quantum states}, Proceedings of the 43rd Annual IEEE Symposium on Foundations
  of Computer Science, pp 429-438.

\bibitem{cleve:ip}
R.~Cleve, Wim van Dam, M.~Nielsen, and A.~Tapp (1998), {\it Quantum entanglement and the communication complexity of the inner
  product function}, Proceedings of the 1st NASA International Conference on
  Quantum Computing and Quantum Communications, Lecture Notes in Computer
  Science, 1509, pp 61-74, Springer-Verlag, quant-ph/9708019.

\bibitem{osborne:gametheory}
M.~Osborne and A.~Rubinstein (1994),
 {\it A course in game theory}, MIT Press.

\bibitem{mayer:imp}
D.~Mayers (1997), {\it Unconditionally secure quantum bit commitment is impossible},
Phys. Rev. Lett, 78, pp 3414-3417.

\bibitem{hklo:bitcomm1}
H.-K. Lo and H.F. Chau (1997), {\it Is quantum bit commitment really possible?}, Phys. Rev. Lett., 78.

\bibitem{hklo:bitcomm2}
H.-K. Lo and H.F. Chau (1998), {\it Why quantum bit commitment and ideal quantum coin tossing are
  impossible}, Physica D, 120.

\bibitem{fannes:ineq}
M.~Fannes (1973), {\it A continuity property of the entropy density of spin lattice systems},
Comm. Math. Phys., 31, pp 291-294.

\end{thebibliography}
\end{document}